\documentclass[,aps,prb,twocolumn,showpacs]{revtex4}
\usepackage{graphicx}
\usepackage{textcomp}
\begin{document}

\title{Thermally stable carbon-related centers in 6H-SiC:\\
  photoluminescence spectra and microscopic models}

\author{Alexander Mattausch}
\email{Mattausch@physik.uni-erlangen.de}

\author{Michel Bockstedte}

\author{Oleg Pankratov}

\affiliation{Theoretische Festk\"orperphysik, Universit\"at
  Erlangen-N\"urnberg, Staudtstr. 7, 91058 Erlangen, Germany}

\author{John W. Steeds}
\email{J.W.Steeds@bristol.ac.uk}

\author{Suzanne Furkert}

\author{Jonathan M. Hayes}

\author{Wayne Sullivan}

\affiliation{Department of Physics, University of Bristol, BS8 1TL Bristol,
  UK}

\author{Nick G. Wright}

\affiliation{Department of Electrical, Electronic \& Computer Engineering, University of Newcastle, UK}

\date{\today}

\begin{abstract}
  Recent \emph{ab initio} calculations [Mattausch \emph{et al.}, Phys. Rev. B
  \textbf{70}, 235211 (2004)] of carbon clusters in SiC reveal a possible
  connection between the tricarbon antisite (C$_{3}$)$_{\text{Si}}$ and the U
  photoluminescence center in 6H-SiC [Evans \emph{et al.}, Phys.~Rev.~B
  \textbf{66}, 35204 (2002)]. Yet, some of the predicted vibrational modes
  were not observed experimentally. Here we report experiments which indeed
  confirm the existence of a low-energy mode for the U-center (as well as for
  the HT3- and HT4-centers with spectral details similar to the U-center). We
  calculated the isotope splitting for the (C$_{3}$)$_{\text{Si}}$-defect and
  found near-perfect agreement with our data. In addition, we discuss the
  carbon di-interstitial (C$_{2}$)$_{\text{Hex}}$ as a model for the Z- and
  HT5-centers. The isotope splitting is also well reproduced, but the absolute
  values of the local mode energies show a discrepancy of about 10\,meV.
\end{abstract}

\pacs{61.80.-x, 63.20.Pw, 78.55.-m}

\bibliographystyle{apsrev}

\maketitle

Silicon carbide (SiC) is a wide band-gap semiconductor especially suitable for
high-power, high-temperature and high-current applications. Due to its extreme
hardness and low diffusion coefficients, the method of choice for the creation
of the doping profiles is ion implantation. However, implantation causes a
severe damage of the crystal and a subsequent annealing at high temperatures
is necessary to eliminate the defects---mostly vacancies and interstitials.
Unfortunately, not all interstitials and vacancies recombine. They can also
cluster and form thermally stable aggregates. One example is the well-known
D$_{\text{II}}$-center,\cite{Pa73} which is most likely an aggregation of
carbon atoms around carbon antisites.\cite{Ma03,Ma04,Ga03a}

Recent photoluminescence (PL) measurements\cite{Ev02} of electron-irradiated
6H-SiC have shown that besides the D$_{\text{II}}$-center other intrinsic
defect centers exist which also possess localized vibrational modes (LVM's)
above the SiC bulk phonon energy. These are the P$-$T centers, which were
tentatively assigned to carbon split-interstitials or di-carbon
antisites,\cite{Ma04,Ga03a} and the U-center, for which a single LVM at nearly
250\,meV was observed. The three-fold isotope splitting in $^{13}$C-enriched
samples signifies a carbon-related defect, although carbon-based defects with
such a high vibrational frequency have not been known in SiC. Yet, \emph{ab
  initio} calculations predict\cite{Ma04} that such defects may exist. Namely
the tri-carbon antisite (C$_{3}$)$_{\text{Si}}$ has several LVM's up to
250\,meV. The energetically highest mode even reproduces the three-fold
isotope splitting of the U-center. However, the missing phonon replicas in
available PL spectra could not be explained in Ref.~\onlinecite{Ma04}. Yet,
the clear identification of observed carbon-related defect centers is
important since it may substantiate the carbon aggregation phenomenon in SiC
predicted by earlier theoretical work.\cite{Ma03,Ma04,Ga03a} Besides, at
elevated temperatures these aggregates can re-emit
carbon-interstitials\cite{Bo04} and thereby facilitate the formation of
further defect centers like D$_{\text{I}}$ or the alphabet-lines.\cite{di}

In this paper we present photoluminescence results for $^{13}$C-enriched
6H-SiC samples and compare them with \emph{ab initio} calculations. The data
show an additional phonon replica of the U-center at 150\,meV, which is in
agreement with the calculated LVM of (C$_{3}$)$_{\text{Si}}$. Further centers,
which are similar to the U-center and labeled HT3 and HT4, were also observed.
With this new experimental data in hand, the centers are well explained by the
tri-carbon antisite (C$_{3}$)$_{\text{Si}}$ model. The centers Z and HT5 are
also discussed. These centers possess two vibrational modes above the SiC bulk
phonon spectrum at about 170\,meV and 200\,meV and also exhibit a three-fold
isotope splitting of the energetically highest mode. Their relationship with
the carbon di-interstitials is discussed. We show that a three-fold isotope
splitting with the frequency ratios of a C$-$C dumbbell (which is the square
root of the mass ratios), does not necessarily originate from a
dumbbell-shaped defect. Both the tri-carbon antisite and the carbon
di-interstitial represent complex defect configurations.

In the following, we first present the experimental details and results and
then turn to the theoretical calculations. Theoretical results are then
compared with the experimental data.

The investigated samples were n- and p-type 6H-SiC with typical doping levels
in the range 10$^{15}-$10$^{16}$\,cm$^{-3}$. They included both epitaxially
grown layers and Lely grown crystals. The p-type specimens showed strong
donor-acceptor pair-luminescence. To produce the defect centers the samples
were electron irradiated with electron energies below and above the silicon
displacement threshold at about 250\,keV. The irradiation was performed using
an ion-free transmission electron microscope. To allow for an identification
of the defects via the isotope splitting of the phonon replicas, samples with
a $^{13}$C enrichment of 30\% were used in addition to the samples with
natural isotope abundance. For the PL measurements the samples were cooled to
7\,K using Oxford Instruments microstats on Renishaw microRaman spectrometers.
The centers were excited using laser wavelengths of 325\,nm and 488\,nm.

\begin{table}
  \caption{\label{tab:centers} ZPL's and LVM's of the centers
    in 6H-SiC.}
  \begin{ruledtabular}
    \begin{tabular}{ccccc}
      center & ZPL (nm) & ZPL (eV) & LVM1 (meV) & LVM2 (meV) \\
      \hline
      Z & 512.5 & 2.417 & 171.6 & 202.0\\
      HT5 & 509.8 & 2.431 & 169.0 & 201.7\\
      HT3 & 492.9 & 2.515 & 151.0 & 246.3\\
      U & 525.0 & 2.361 & 151.6 & 246.6\\
      HT4 & 546.4 & 2.268 & 149.9 & 244.7
    \end{tabular}
  \end{ruledtabular}
\end{table}

The reported centers were produced by high-temperature anneals in the range
1000$-$1500$^{\circ}$C. The U- and the Z-center came up at the lower end of
this range and persisted in some cases up to 1500$^{\circ}$C. The centers HT3
and HT4 were observed after 1200$^{\circ}$C and 1300$^{\circ}$C anneals,
respectively. While HT3 was always present, HT4 appeared only in some samples.
The HT5-center was observed after the highest temperature of 1500$^{\circ}$C.

All of the optical transitions were excited by both the 325\,nm and the
488\,nm wavelength. A major effort was put into the deconvolution of the
observed spectra, since considerable overlap occurred between the spectral
details. Employing three different techniques we were able to unambiguously
assign the spectral details to a specific zero phonon line (ZPL). One method
was by spatial separation. Spectra were obtained at a series of points along
lines through the irradiated region, and the relative intensities of the ZPL's
changed with position. A second approach was the comparison of the results
after different annealing temperatures, which also affected the relative
intensities of the spectra. The third method was the investigation of a wide
range of specimens irradiated with different electron doses and energies.
Further details on this approach can be found in Refs.~\onlinecite{Ev02}
and~\onlinecite{St02}.

\begin{figure*}
  \centering
  \includegraphics[width=\linewidth]{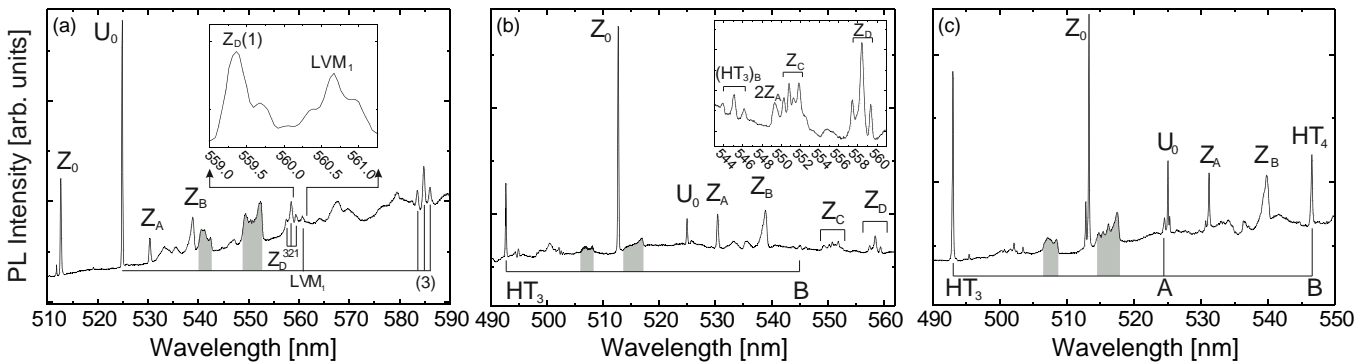}
  \caption{\label{fig:Exp} Spectra of 6H-SiC samples (taken at $T=7$\,K with a
    laser wavelength of 325\,nm) with a $^{13}$C-enrichment of 30\% and
    natural isotope abundance showing the U-, Z-, HT3- and HT4-centers. The
    shaded regions indicate the bulk phonons belonging to the marked ZPL. (a)
    The U- and the Z-center in a $^{13}$C-enriched sample. The inset displays
    the broadened lower-energy mode of the U-center and the
    $^{12}$C$-^{12}$C-peak of the energetically highest Z-mode. (b) The HT3-
    and the Z-center in a $^{13}$C-enriched sample. All modes of the Z-center
    are visible. The inset displays the isotope splitting of the modes
    Z$_{\text{D}}$ and Z$_{\text{C}}$ as well as the three-fold splitting of
    the highest HT3-mode. (c) The centers Z, U, HT3 and HT4 in a sample with
    natural isotope abundance. The overlap of the ZPL of HT4 and the
    high-energy mode of HT3 is clearly visible, as well as the narrow
    separation between the lower-energy mode of HT3 and the ZPL of the
    U-center.}
\end{figure*}

Figure~\ref{fig:Exp}a shows the PL spectra of the U- and the Z-centers in
$^{13}$C-enriched 6H-SiC. Both centers were originally discussed in
Ref.~\onlinecite{Ev02}. The three-fold isotope splitting of the high-energy
mode at 246.6\,meV is clearly visible. In addition to the single originally
observed mode\cite{Ev02} a further mode with a lower energy was observed at
151.6\,meV in samples with natural C abundance (cf.\ Table~\ref{tab:centers}).
In isotope-enriched SiC this mode splits in a complex fashion (cf.\
Fig.~\ref{fig:Exp}a). The centers HT3 and HT4 possess properties similar to
those of the U-center. Figure~\ref{fig:Exp}b shows the HT3-center alongside
the Z-center in $^{13}$C-enriched 6H-SiC. Each center also has a high-energy
mode around 245\,meV (cf.\ Table~\ref{tab:centers}) exhibiting a three-fold
isotope splitting (cf.\ inset of Fig.~\ref{fig:Exp}b for HT3) and a
lower-energy mode around 150\,meV. Some aspects of the HT3- and HT4-centers
deserve particular mention. First, in the sample with natural C-abundance the
high-energy mode coincides with the ZPL of HT4 at 546\,nm (cf.\
Fig.~\ref{fig:Exp}c). Clear evidence of a LVM at this wavelength could be
deduced from a $^{13}$C-enriched sample which lacks the HT4-luminescence and
shows a clear three-fold isotope splitting. The second aspect is the small
separation between the lower-energy mode of HT3 and the ZPL of the U-center
(cf.\ Fig.~\ref{fig:Exp}c). Although the LVM could be clearly identified in
the sample with natural C-abundance, the isotope splitting could not be
deduced.

The Z-center is clearly present in all shown spectra. Its full details are
visible in Fig.~\ref{fig:Exp}b. It possesses 4 LVM's, yet only two of them
(Z$_{\text{C}}$ and Z$_{\text{D}}$ in Fig.~\ref{fig:Exp}b) are above the SiC
bulk phonon spectrum at 172\,meV and 202\,meV. The HT5-center possesses
similar properties (cf.\ Table~\ref{tab:centers}). In $^{13}$C-enriched
samples the high-energy mode shows a three-fold isotope splitting, while the
lower-energy mode exhibits a complex broadening (cf.\ Fig.~\ref{fig:Exp}b).

The high vibrational energy and the three-fold splitting of the highest mode
upon carbon isotope substitution suggest that the observed defect centers are
carbon-related. Hence we focused our theoretical modeling on
carbon-interstitial and carbon-antisite complexes. We employed
density-functional theory with the local density approximation (LDA) as
implemented in the software package FHI96SPIN.\cite{Bo97} Supercells with 216
sites for 3C-SiC and 128 sites for 4H-SiC and a plane wave basis set with a
cut-off energy of 30\,Ry were used. The Brillouin zone was sampled by a single
$\Gamma$-point for the large 3C-SiC supercells. For 4H-SiC a $2 \times 2
\times 2$ Monkhorst-Pack mesh\cite{Mo76} was used except for the calculation
of the LVM's, where we used only the $\Gamma$-point. Due to the large
supercells the defect molecule approximation was employed for the analysis of
the LVM's. In this approach the calculation of the dynamical matrix is
restricted to the defect and its nearest-neighbor atoms (cf.
Ref.~\onlinecite{Ma04} for details). We verified that this approximation
affects only the lowest-energy LVM's close to the bulk phonon band.\cite{Ma03}

\begin{figure}
  \centering
  \includegraphics[width=0.8\linewidth]{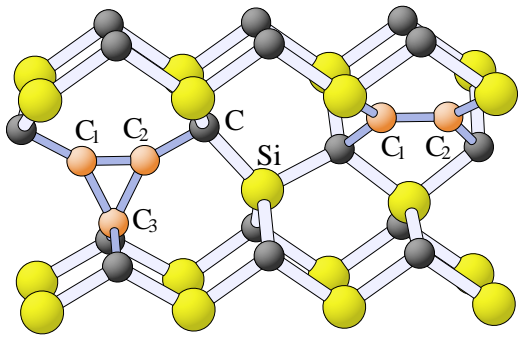}
  \caption{The tri-carbon antisite (C$_{3}$)$_{\text{Si}}$ (left) and the carbon
    di-interstitial (C$_{2}$)$_{\text{Hex}}$ (right) in 3C-SiC.}
  \label{fig:models}
\end{figure}

The candidates for the centers described above must possess two properties:
they must exhibit a three-fold isotope splitting of the highest mode in
$^{13}$C-enriched material and they must be thermally stable. Two
possibilities are the carbon di-interstitial (C$_{2}$)$_{\text{Hex}}$ and its
structurally similar configurations in 4H-SiC and the tri-carbon antisite
(C$_{3}$)$_{\text{Si}}$. The calculations were performed for 3C- and 4H-SiC. A
close agreement of the results for the two polytypes ensures the
transferability to the more complex 6H-SiC. The defects are depicted in
Fig.~\ref{fig:models} for 3C-SiC. The thermal stability is guaranteed by high
dissociation energies, which are 4.8\,eV (5.8\,eV) for (C$_{3}$)$_{\text{Si}}$
in 3C-SiC (4H-SiC) and 4.8\,eV for (C$_{2}$)$_{\text{Hex}}$ in 3C-SiC. In
4H-SiC, various configurations exist for the di-interstitial with dissociation
energies between 5.1\,eV and 5.5\,eV. The most stable defect is
(C$_{\text{sp}}$)$_{2,\text{hh}}$ with two carbon split-interstitials at
neighbouring sites. The interstitial carbon atoms relax towards each other, so
that a configuration similar to (C$_{2}$)$_{\text{Hex}}$ in 3C-SiC forms. All
these defects are neutral for practically all positions of the Fermi-level.

\begin{table*}
  \caption{\label{tab:lvms} Calculated LVM's in meV of the carbon
    di-interstitials (C$_{2}$)$_{\text{Hex}}$ and the tri-carbon antisite
    (C$_{3}$)$_{\text{Si}}$. In 4H-SiC, further di-interstitials exist with
    similar geometrical configuration and similar vibrational properties 
    (see Ref.~\onlinecite{Ma04}). The letters $g$ and $u$ denote symmetric and
    antisymmetric vibrations, respectively.}
  \begin{ruledtabular}
    \begin{tabular}{ccccccc}
      LVM & (C$_{2}$)$_{\text{Hex}}$ (3C) &
      (C$_{\text{sp}}$)$_{2,\text{hk,cub}}$ (4H) &
      (C$_{\text{sp}}$)$_{2,\text{hh}}$ (4H) & (C$_{3}$)$_{\text{Si}}$ (3C) &
      (C$_{3}$)$_{\text{Si,k}}$ (4H) & (C$_{3}$)$_{\text{Si,h}}$ (4H) \\
      \hline
      1 & 132.5 (u) & 128.0 (u) & 134.9 (u) & 118.6 (u)& 119.0 (u) & 119.0 (u)\\
      2 & 160.4 (g) & 161.3 (g) & 159.9 (g) & 129.8 (u) & 130.2 (u) & 130.5 (u)\\
      3 & 167.3 (u) & 167.1 (u) & 168.1 (u) & 148.7 (g) & 154.0 (g) & 154.2 (g)\\
      4 & 184.3 (g) & 189.1 (g) & 191.5 (g) & 181.3 (u) & 182.3 (u) & 182.6 (u)\\
      5 &           &           &           & 248.5 (g) & 254.9 (g) & 255.7 (g)\\
    \end{tabular}
  \end{ruledtabular}
\end{table*}

\begin{figure*}
  \centering
  \includegraphics[width=\linewidth]{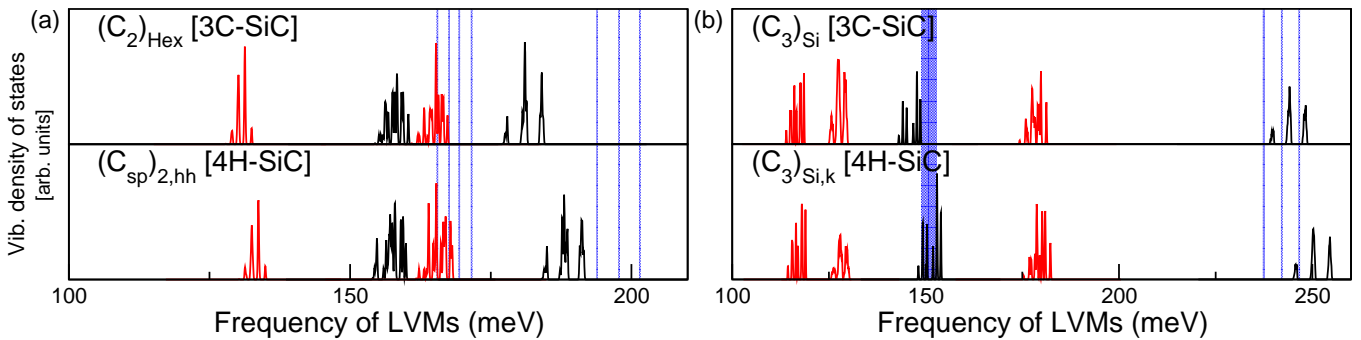}
  \caption{\label{fig:LVM-model} (Color online) Calculated LVM's of (a) the
    carbon di-interstitials and (b) the tri-carbon antisite in 3C and 4H-SiC
    with a $^{13}$C-enrichment of 30\%. The black peaks indicate symmetric
    modes, the red (gray) peaks are antisymmetric vibrations. The phonon
    replicas of the Z-center (a) and the U-center (b) including isotope shifts
    are indicated by the blue (shaded) lines.}
\end{figure*}

Both defects show a distinct vibrational pattern. The frequencies of the LVM's
in 3C- and 4H-SiC are listed in Table~\ref{tab:lvms}. Since the defects
possess mirror symmetry (which is in 4H-SiC limited to the defect molecule),
the modes are marked as symmetric ($g$) or antisymmetric ($u$) vibrations.
This is important for PL experiments, since the bound exciton mostly interacts
with a symmetric but not with an antisymmetric vibration.\cite{Pi86} It is
clearly visible from Table~\ref{tab:lvms} that the LVM's of the defects are
practically independent of the polytype and the involved cubic or hexagonal
sites. We therefore expect that our results for the 3C and 4H polytypes are
transferrable to 6H-SiC.

The di-interstitial possesses four LVM's above the SiC bulk phonon spectrum.
The LVM's at 160\,mev and 190\,meV are symmetric vibrations (cf.
Table~\ref{tab:lvms}). The highest mode shows a clear three-fold isotope
splitting, while the lower-energy modes show a complex broadening or splitting
(cf.\ Fig.~\ref{fig:LVM-model}a). The relative height of the peaks within a
mode results from the various possibilities of substituting a defect's atom
with $^{13}$C assuming an enrichment of 30\%. The peak heights of different
modes are independent and cannot be compared. Considering the two symmetric
modes of (C$_{2}$)$_{\text{Hex}}$, the spectrum is similar to that of the Z-
and the HT5-center. Besides, as in 4H-SiC various different defect
configurations involving the two cubic and the hexagonal sites are possible in
6H-SiC, allowing for different exciton binding energies (and therefore
different ZPL's) with comparable phonon replicas. However, the calculated
values are systematically about 10\,meV lower than the measured frequencies.
This finding is especially noteworthy since the LDA tends to overestimate the
calculated frequencies.

The tricarbon antisite (C$_{3}$)$_{\text{Si}}$ has five LVM's above the bulk
phonon spectrum. Symmetric modes are the mode 5 at 250\,meV and the mode 3 at
150\,meV. All other modes are antisymmetric (cf.\ Table~\ref{tab:lvms}). The
isotope splitting of the defect in 3C- and 4H-SiC is displayed in
Fig.~\ref{fig:LVM-model}b alongside the measured frequencies for the U-center
(the centers HT3 and HT4 have similar phonon replicas, cf.\
Table~\ref{tab:centers}). The mode 5 shows a clear three-fold isotope
splitting as one would expect for a simple carbon dumbbell vibration. This
simple pattern results from the localization of the vibration onto the atoms
C$_{1}$ and C$_{2}$ of the triangular defect, whereas the neighboring atoms
are practically not involved. In contrast, the neighbor atoms do participate
in the vibration of the mode 3. Consequently, this mode exhibits a significant
broadening due to the $^{13}$C-enrichment. The measured values of the U-,
HT3-, and HT4-centers are in excellent agreement with the calculated absolute
values of the symmetric modes of (C$_{3}$)$_{\text{Si}}$. The isotope
splitting of the centers is also well reproduced. The antisymmetric modes
should be suppressed in photoluminescence as discussed above. In 6H-SiC the
defect can be located at the cubic sites $k_{1}$ and $k_{2}$ or at the
hexagonal site $h$, providing three different defect configurations. As
observed for 4H-SiC, the difference of the vibrational patterns for the
different sites shall be negligible. Thus we believe that
(C$_{3}$)$_{\text{Si}}$ should be a good model for the centers U, HT3 and HT4
in 6H-SiC.

In conclusion, we presented thermally stable photoluminescence centers in
6H-SiC with localized vibrational modes up to 250\,meV. All centers show a
clear three-fold isotope splitting of the highest mode with the frequency
ratio typical for a carbon-carbon dumbbell vibration. Based on the vibrational
spectra obtained from \emph{ab initio} calculations, the di-interstitial
(C$_{2}$)$_{\text{Hex}}$ is considered as a model for the centers Z and HT5,
although a systematic frequency shift of 10\,meV is observed. The results
obtained for the tri-carbon antisite (C$_{3}$)$_{\text{Si}}$ are in excellent
agreement with the U-, HT3-, and HT4-centers. This identification demonstrates
the possibility of carbon aggregates predicted earlier
theoretically.\cite{Ma03,Ma04,Ga03a}

  This work was supported by the Deutsche Forschungsgemeinschaft
  within the SiC Research Group and the UK Engineering and Physical Science
  Research Council.

\end{document}